\documentclass[aip,reprint,amsmath,amssymb,prapplied,superscriptaddress]{revtex4-1}

\usepackage{graphicx,url,float}
\usepackage[breaklinks=true,colorlinks=true,linkcolor=blue,urlcolor=blue,citecolor=blue]{hyperref}
\usepackage{dcolumn}
\usepackage{bm}
\newcommand{\uS}[0]{\,$\mu$s}

\begin{document}

\title{Network of sensitive magnetometers for urban studies}

\author{T. A. Bowen}
\email{tbowen@berkeley.edu}
\affiliation{Department of Physics, University of California, Berkeley, California 94720-7300, USA}
\affiliation{Space Sciences Laboratory, University of California, Berkeley, California 94720-7300, USA}\
\author{E. Zhivun}
\affiliation{Department of Physics, University of California, Berkeley, California 94720-7300, USA}
\author{A. Wickenbrock}
\affiliation{Johannes Gutenberg-Universit\"{a}t Mainz, 55128 Mainz, Germany}
\author{V. Dumont}
\affiliation{Department of Physics, University of California, Berkeley, California 94720-7300, USA}
\author{S. D. Bale}
\affiliation{Department of Physics, University of California, Berkeley, California 94720-7300, USA}
\affiliation{Space Sciences Laboratory, University of California, Berkeley, California 94720-7300, USA}\
\author{C. Pankow}
\affiliation{Center for Interdisciplinary Exploration \& Research in Astrophysics (CIERA), Northwestern University, Evanston, IL 60208, USA}
\author{G. Dobler}
\affiliation{Center for Urban Science and Progress, New York University, Brooklyn, NY 11201, USA}
\author{J. S. Wurtele}
\affiliation{Department of Physics, University of California, Berkeley, California 94720-7300, USA}
\author{D. Budker}
\affiliation{Department of Physics, University of California, Berkeley, California 94720-7300, USA}
\affiliation{Johannes Gutenberg-Universit\"{a}t Mainz, 55128 Mainz, Germany}
\affiliation{Helmholtz Institut Mainz, 55128 Mainz, Germany}
\affiliation{Lawrence Berkeley National Laboratory, Berkeley, CA 94720}

\date{\today}

\begin{abstract}
The magnetic signature of an urban environment is investigated using a geographically distributed network of fluxgate magnetometers deployed in and around Berkeley, California. The system hardware and software are described and initial operations of the network are reported. The sensors measure vector magnetic fields at a 3960\,Hz sample rate and are sensitive to fluctuations below 0.1\,$\textrm{nT}/\sqrt{\textrm{Hz}}$. Data from geographically separated stations are synchronized to $\pm100$\uS{} using GPS and computer system clocks and automatically uploaded to a central server. Observations of several common sources of urban magnetic fields are reported, as well as a transient event identified as a lightning strike. A wavelet analysis is used to study observations of the urban magnetic field over a wide range of temporal scales. The Bay Area Rapid Transit (BART) is identified as dominant signal in our observations, exhibiting significant differences in both day/night and weekend/weekday signatures. A superposed epoch analysis is used to study and extract the BART signal. We furthermore present initial results of the correlation between sensors.
\end{abstract}

\maketitle

\section{Introduction}

The study of fluctuating magnetic fields has found widespread application in a variety of disciplines. High-cadence (or effective sampling rate) magnetic field measurements have been used for magnetic-anomaly detection (MAD), which has numerous security applications, such as naval defense and unexploded ordnance detection \cite{sheinker:2009}. Additionally, geographically distributed magnetometers, operating at high sample rates, have been designed to study the global distribution of lightning strikes using the Schumann resonance \cite{Schlegel:2002}.\\

Low-frequency magnetic field measurements, typically corresponding to large length scales, provide important information relating to the nature of magnetic sources in the Earth's core, maps of near-surface fields, as well as crustal composition and structure models \cite{egbert:1997}. An international consortium of magnetometer arrays, known as SuperMag, comprises approximately 300 magnetometers operated by numerous organizations \cite{Gjerloev:2009,Gjerloev:2012}.  The magnetic field data collected from these stations is uniformly processed and provided to users in a common coordinate system and timebase \cite{Gjerloev:2012}. Such data are important to global-positioning-system (GPS)-free navigation, radiation-hazard prediction \footnote{\url{http://www.nws.noaa.gov/os/assessments/pdfs/SWstorms_assessment.pdf}}, climate and weather modeling, and fundamental geophysics research. Additionally, measurements of the auroral magnetic field are necessary in testing models for space-weather prediction, which aims to mitigate hazards to satellite communications and GPS from solar storms \cite{Angelopoulos:2008,Peticolas:2008,Harris:2008}.\\

Magnetometry has additionally been applied in the search for earthquake precursors. Anomalous enhancements in ultralow frequency (ULF) magnetic fields were reported leading up to the October 17, 1989 Loma Prieta earthquake \cite{fraser-smith:1990}. Similar anomalous geomagnetic behavior was observed for the month preceding the 1999 Chi-Chi earthquake in Taiwan \cite{yen:2004}. Recently, attempts have been made to study precursors to the 2011 Tohoku earthquake in Japan \cite{hayakawa:2015}.\\

\begin{figure*}[ht]
  \centering
  \includegraphics[width=0.8\textwidth]{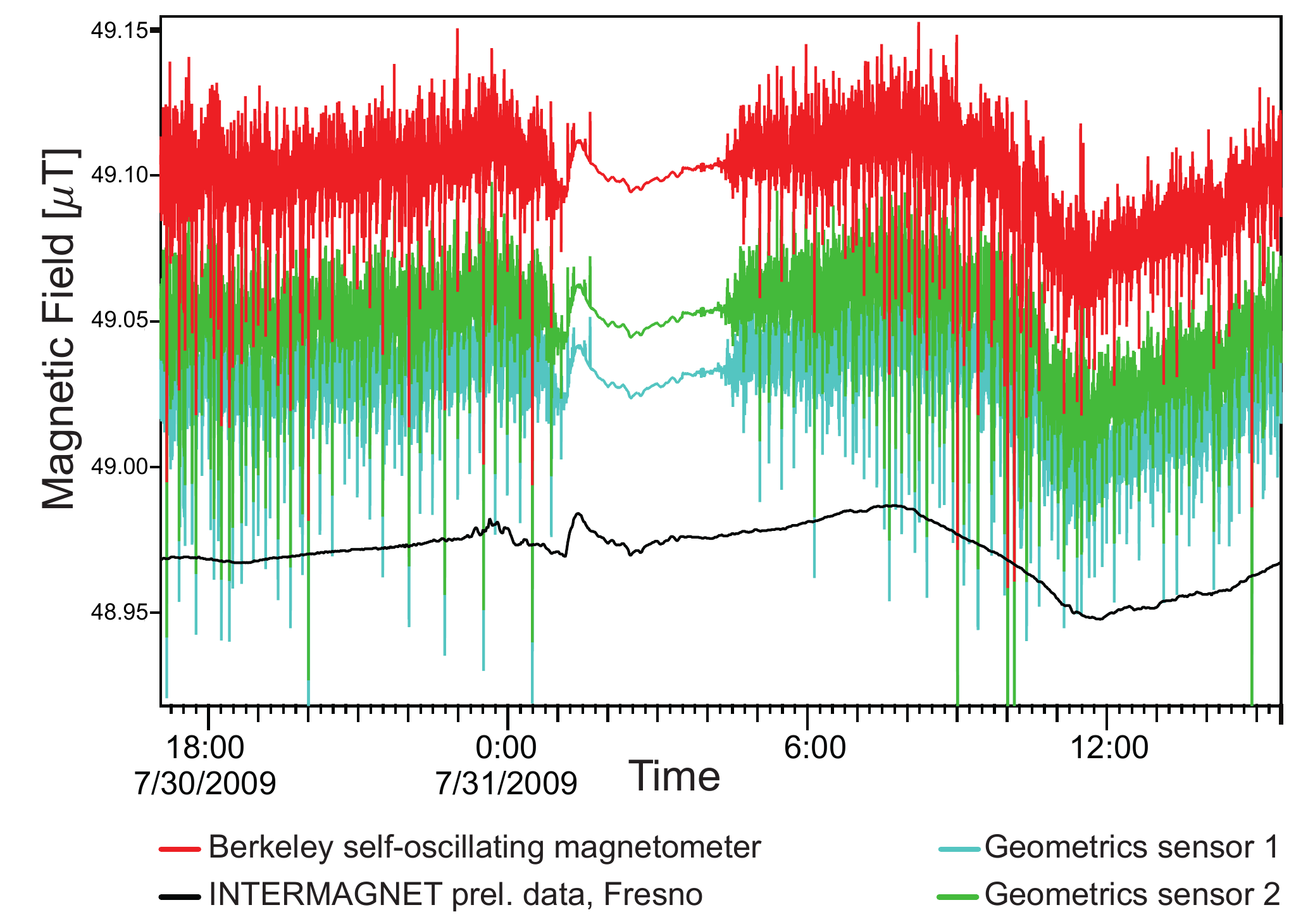}
  \caption{The Earth's magnetic field, as recorded at the UC Berkeley Botanical Garden.  The red trace represents measurements taken by an all-optical self-oscillating magnetometer designed and constructed at UC Berkeley in partnership with Southwest Sciences, Inc. \cite{Higbie:2006}.  The green and blue traces are data taken with a commercially available cesium magnetometer from Geometrics, Inc.  The gray trace represents the average field seen by a geomagnetic observatory located in Fresno, CA.}
  \label{berkeley_garden}
\end{figure*}

Despite its numerous applications, magnetometry is frequently limited by contamination from unrelated, and often unknown, sources. In their search for earthquake precursors, Fraser-Smith et al. (1978) found that magnetic noise from the Bay Area Rapid Transit (BART) system dominated their ULF sensors \cite{fraser-smith:1978}. This noise is evident in Fig. \ref{berkeley_garden}, which depicts the magnitude of the magnetic field recorded at the University of California (UC) Berkeley Botanical Garden. Despite the obvious presence of local permanent or time-varying magnetic contamination, the recorded magnetic field is dominated by fluctuations beyond the expected typical geomagnetic values. These fluctuations diminish dramatically between approximately 1 AM and 5 AM local time, indicating that these are the same fluctuations that are attributed to the operation of BART \cite{fraser-smith:1978}.\\

Even during the magnetically quieter nighttime period, there remain fluctuations which exceed expected geophysical values. Certainly, some of the field variation can be attributed to variations in the ionospheric dynamo and other natural sources: similar trends appear in the magnetic record from the Botanical Garden as well as Intermagnet data from a magnetometer located in the nearby city of Fresno \cite{chapman:1962,kerridge:2001}. Nevertheless, the contributions of human activity to these nighttime fluctuations remain poorly understood.\\

Only recently have investigators been able to accumulate and study the increasing quantity of spatially and temporally granular data that characterize the evolving state of a city. These data include information from social networks (e.g., twitter), financial transactions, transportation systems, environmental markers (pollution, temperature, etc.) and a wide range of other physical quantities. Passive observations of cities not only serve as a means of quantifying urban functioning for the purpose of characterizing cities as complex systems of study, but they can also yield tremendous benefit to city agencies and policy makers.\\

Recent work has shown that broadband visible observations at night can identify patterns of light that can be used to measure total energy consumption and public health effects of light pollution on circadian rhythms \cite{dobler:2015}. High frequency (${\sim}120$\,Hz) measurements of urban lighting allow for phase change detection of fluorescent and incandescent flicker, which may be used as a predictor for power outages \cite{bianco:2016}. In addition, infrared hyperspectral observations can determine the molecular content of pollution plumes produced by building energy consumption providing a powerful method for environmental monitoring of cities \cite{ghandehari:2017}.\\

Here we report on the development of a synchronized magnetometer array in Berkeley, California for the purpose of studying the signature of urban magnetic field over a range of spatiotemporal scales. The array, currently consisting of four magnetometers operating at a 3960 Hz sample rate, will make sustained and continued measurements of the urban field over years, observing the city's dynamic magnetic signature. Through systematically observing magnetic signatures of a city, we hope to complement advances made elsewhere in urban informatics and applied physics to provide a deeper understanding of urban magnetic noise for researchers in geophysics and earth science.\\
 
In Sec.~\ref{development} we briefly describe the components and performance of the hardware and software implemented in our magnetometer array. We emphasize our preference for commercially available hardware and advanced timing algorithms and utilize  techniques  similar to those of the Global Network of Optical Magnetometers for Exotic physics searches (GNOME) project \cite{pustelny:2013}.\\

 The analysis of our data is presented in Sec.~\ref{signatures}. In Sec.~\ref{singleobservations} we  present the signatures of several common urban sources and observe the signature of a lightning strike. Station data are analyzed and compared in Sec.~\ref{multistation}. Clear variations between weekday and weekend, day and night, and distance from BART tracks are observed. Measurements are examined with wavelets in Sec.~\ref{correlation}. Initial results of correlating station data are presented. In Sec.~\ref{extraction}, we present an initial method to isolate the BART signal. Conclusions and directions for future research are presented in Sec.~\ref{discussion}.

\section{Instrumentation, Hardware, and Data Acquisition}
\label{development}

\subsection{Description of the system}

\begin{figure*}[ht]
  \centering
  \includegraphics[width=\textwidth]{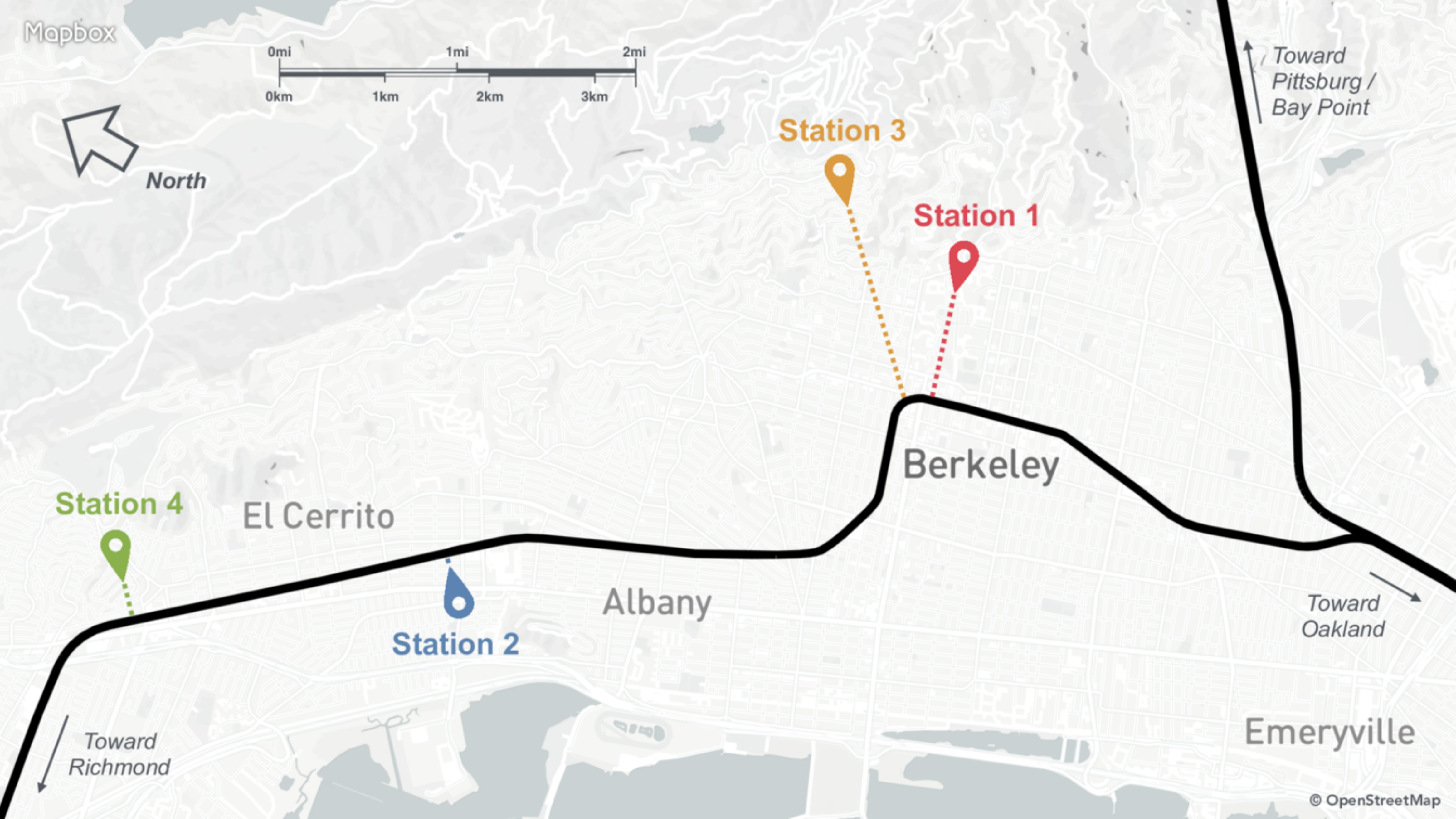}
  \caption{Map of Berkeley and location of the stations. The colored pins identify the locations of the magnetometers in our network. The black line shows the different paths of the BART trains. The shortest distance from each magnetometer station to the nearby BART line is represented by the dashed lines. Stations 1 to 4 are respectively located 1000, 130, 2000 and 360 meters from the closest BART line.}
  \label{map}
\end{figure*}

Our magnetometer network consists of several spatially separated magnetometers with high-precision timing for correlation analysis. Figure \ref{map} shows a map of Berkeley with the sensor locations of the network. Each station consists of a commercially available fluxgate magnetometer, a general-purpose laptop computer, and an inexpensive GPS receiver. Our approach avoids bulky and expensive hardware, favoring consumer components wherever possible. As a result, we reduce the cost of the acquisition system and enable portability through battery operation. However, achieving the desired timing precision (${\sim}100$\,\uS{}) with affordable commercial hardware requires implementing a customized timing algorithm.\\

\begin{figure}[h]
  \centering
  \includegraphics[width=\columnwidth]{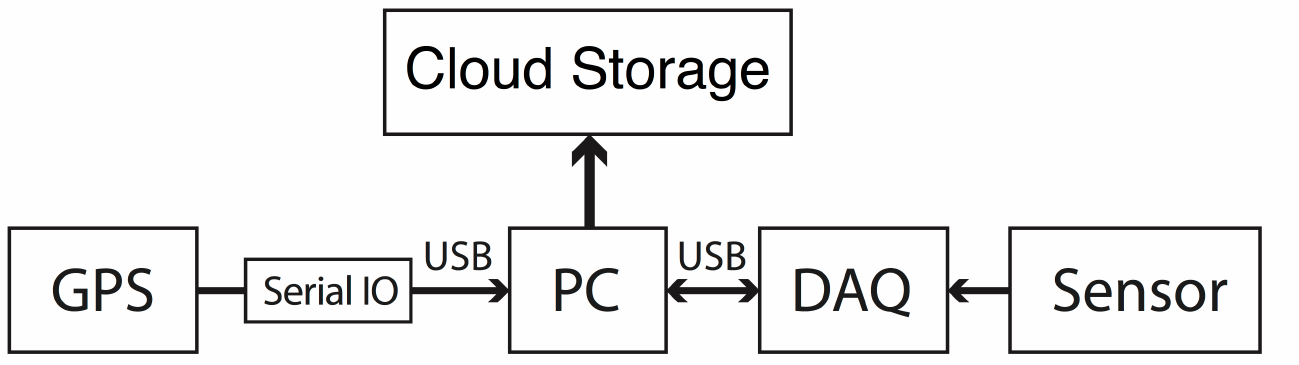}
  \caption{A schematic of a magnetometer station. The computer (PC) retrieves the magnetic field vector value continuously from the data acquisition system (DAQ) that reads out the voltage on the sensor. The timing data are provided by a GPS receiver, with a dedicated 1 pulse-per-second (PPS) output through a Serial IO RS232-to-USB converter. The acquired data are uploaded to a shared Google Drive folder.}
  \label{exp_setup}
\end{figure}

A schematic of a single magnetometer is presented in Fig. \ref{exp_setup}. Each setup is controlled by a computer (PC, ASUS X200M) running the Windows 10 operating system (OS). The PC acquires magnetic field data from the DAQ (Biomed eMains Universal Serial Bus, USB, 24\,bit) and timing data from the GPS receiver (Garmin 18x LVC). The DAQ continuously samples the fluxgate magnetometer (MAG, Biomed eFM3A) at a rate of 3960 sample/s. Absolute timing data are provided once per second by the GPS receiver, which is connected through a powered high-speed RS232-to-USB converter (SIO-U232-59). The GPS pulse-per-second signal, with 1\uS{} accuracy, is routed to the computer through the carrier detect (CD) pin of the RS232 converter. Data from the DAQ arrive in packets of 138 vector samples approximately every 35\,ms. As the data are received, they are recorded together with the GPS information and the computer system clock. Data are uploaded via wireless internet to a shared Google Drive folder.

\subsection{Time synchronization and filtering}
\label{timing}

Time intervals between the GPS updates are measured by the computer system clock (performance counter), which runs at 2533200\,Hz. A linear fit model is used to determine the absolute system time relative to the GPS. Only the GPS timing data from the last 120 seconds are used to determine linear fit parameters. When a magnetic-field data packet is received, the acquisition system immediately records the performance counter value. The packet time-tag is determined from interpolating the linear fit GPS time to the performance counter value. Typical jitter of the inferred time stamps is 120\uS{} and is limited by the USB latency \cite{Korver:2003}.\\

Some of the data packets cannot be processed immediately due to OS limitations. These packets are delayed by up to several milliseconds before being delivered to the data-acquisition software. The number of the delayed packets depends on the system load. During normal operation of a magnetometer, the average fraction of the delayed packets is about $3\%$. The intervals between both GPS and packet arrival times are measured with the performance counter in order to identify the delayed packets. Any GPS data that arrive more than 50\uS{} late are discarded from the linear-fit model. When magnetometer data packet arrives with a 200\uS{} or greater discrepancy from the expected time, their time stamp is replaced with the expected arrival time, which is inferred from the linear fit. Our time filtering algorithm and data acquisition software are publicly available on GitHub \footnote{\url{https://github.com/lenazh/UrbanMagnetometer}}.

\subsection{Performance characterization}

\begin{figure*}[ht]
  \centering
  \includegraphics[width=\textwidth]{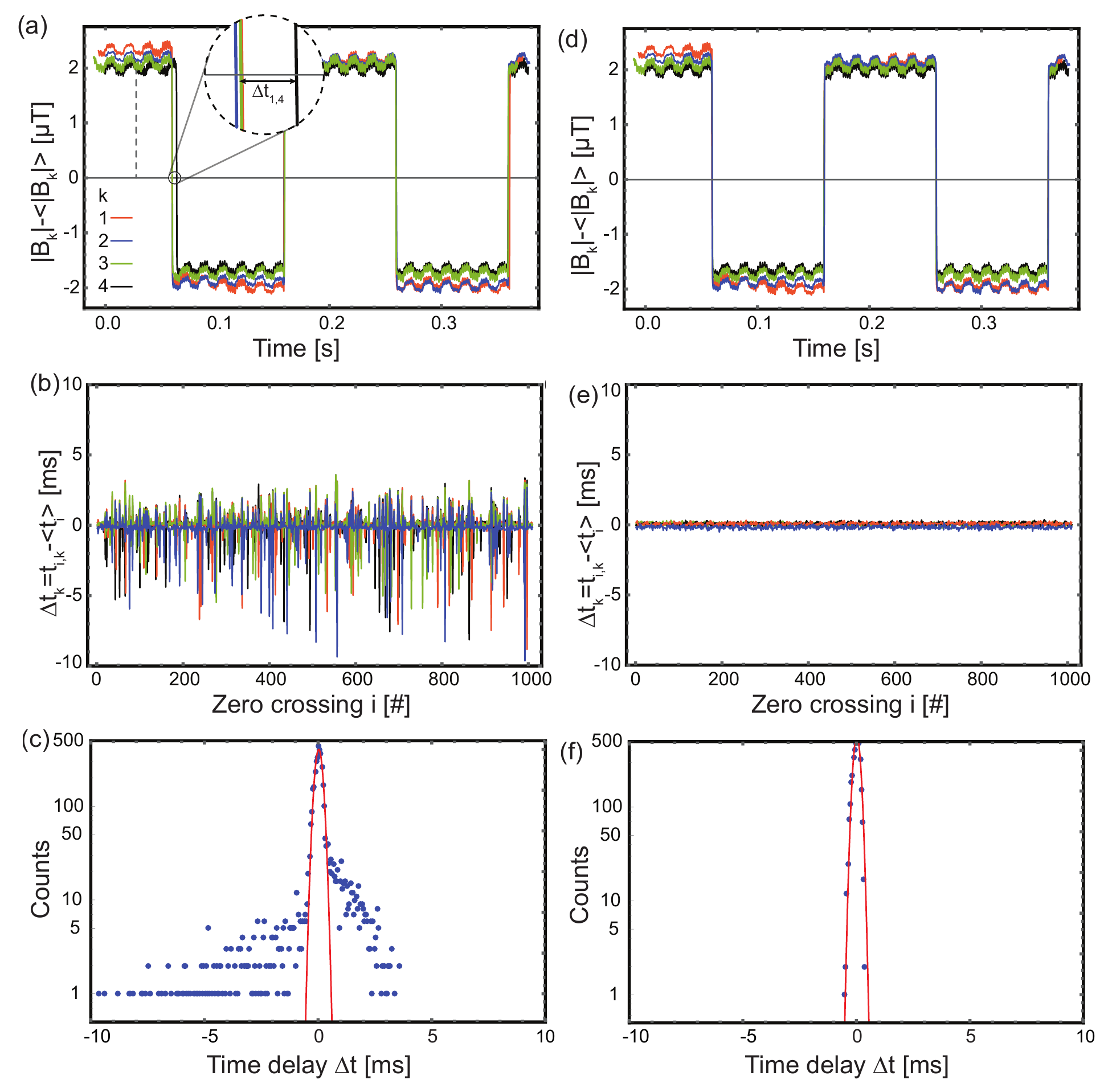}
  \caption{Characterization measurement of the time synchronization algorithm. The four magnetometers are subjected to the same square-wave-modulated magnetic field. The left column shows the measurement with the unprocessed time information. The right column shows the same data with the corrected time. (a) Time series of the four magnetometer traces. The inset shows a time discrepancy of magnetometer 4 at the zero crossing. (b) The difference of the mean zero crossing time of the square wave and the individual zero-crossing time for each magnetometer for two minutes of data. The data show a spread and delays of up to 10\,ms. (c) Histogram of the data in (b), with the red curve representing the best-fit Gaussian. While most zero-crossing events have the correct timing within 120\uS{} (standard deviation of the Gaussian), there are a significant number of outliers. (d), (e) and (f) show the same data after implementing the time synchronization algorithm.}
\label{Figure3}
\end{figure*}

In order to characterize the performance of the data acquisition system, we apply a reference signal simultaneously to the sensors. All four sensors are placed together into a single Helmholtz coil system driven by a pulse generator. The amplitude of the pulses is 2\,$\mu$T, the period is $200$\,ms, and the duty cycle is $50\%$ (Fig. \ref{Figure3}a). The top and bottom rows in Fig. \ref{Figure3} represent the data before and after application of the timing-correction algorithm. The inset in Fig. \ref{Figure3}a demonstrates how a delay in retrieving a data packet disrupts the timing of the field pulses. When a data packet is delayed, the magnetic-field samples are distributed over a slightly larger time period, which affects the estimated time of the field change. In Fig. \ref{Figure3}, interpolated time of the falling edge from sensor four has been several milliseconds after the field changed, causing the zero crossing of the square wave to shift in time from the other sensors. Figure \ref{Figure3}b shows the time discrepancies between the interpolated and expected square wave zero crossings. The zero crossings are recorded with 120\uS{} standard deviation from the expected interval; however, there are a large number of outliers with up to a 10 ms discrepancy. Figure \ref{Figure3}c shows the histogram of the data from \ref{Figure3}b, where the red curve represents the best fit Gaussian. After the timing correction algorithm is applied to the data, time stamps associated with outlier packets are replaced with the expected arrival times (Fig. \ref{Figure3} d,e,f). The remaining jitter is a Gaussian distributed with an error of 120\uS{}.

\subsection{Instrumental noise floor}

\begin{figure*}[ht]
  \centering
  \includegraphics[width=0.9\textwidth]{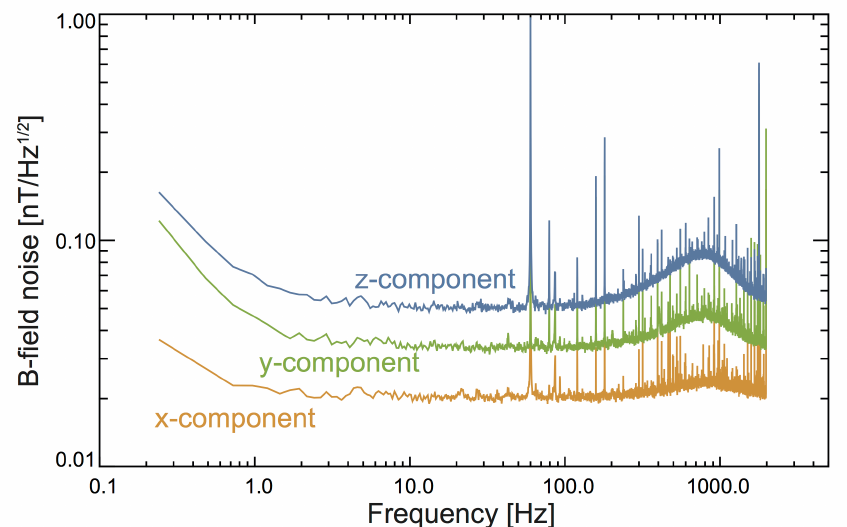}
  \caption{Instrumental noise floors for each vector axis of a Biomed magnetometer.}
  \label{noisefloor}
\end{figure*}

Figure \ref{noisefloor} shows the instrumental noise floors for each vector axis of a Biomed magnetometer. Data were obtained in a two-layer $\mu$-metal shield for approximately 35 minutes ($2^{23}$ samples at 3960 sample/s). Data were separated into an ensemble of 64 individual intervals of $2^{17}$ samples. Power spectral densities were calculated for each interval, with the noise floor taken as the ensemble average of the interval spectra. The noise floor varies between individual axes, with the most noise observed on the Z-axis. For all three directions, the noise floor is constant between $\approx$ 2\,Hz and $\approx$ 700\,Hz. Narrowband spectral features from 60\,Hz and harmonics are easily observed in the data. For frequencies above 1\,Hz, the noise floor is uniformly below 0.1 $\textrm{nT}/\sqrt{\textrm{Hz}}$. The peak observed in the noise floor, predominately in the Z and Y channels, between $\sim$ 300 and $\sim$1500\,Hz, is likely due to the operation of the fluxgate electronics.

\section{Data Analysis}
\label{signatures}

\subsection{Observations of urban magnetic signatures}
\label{singleobservations}

\begin{figure*}[ht]
  \centering
  \includegraphics[width=\textwidth]{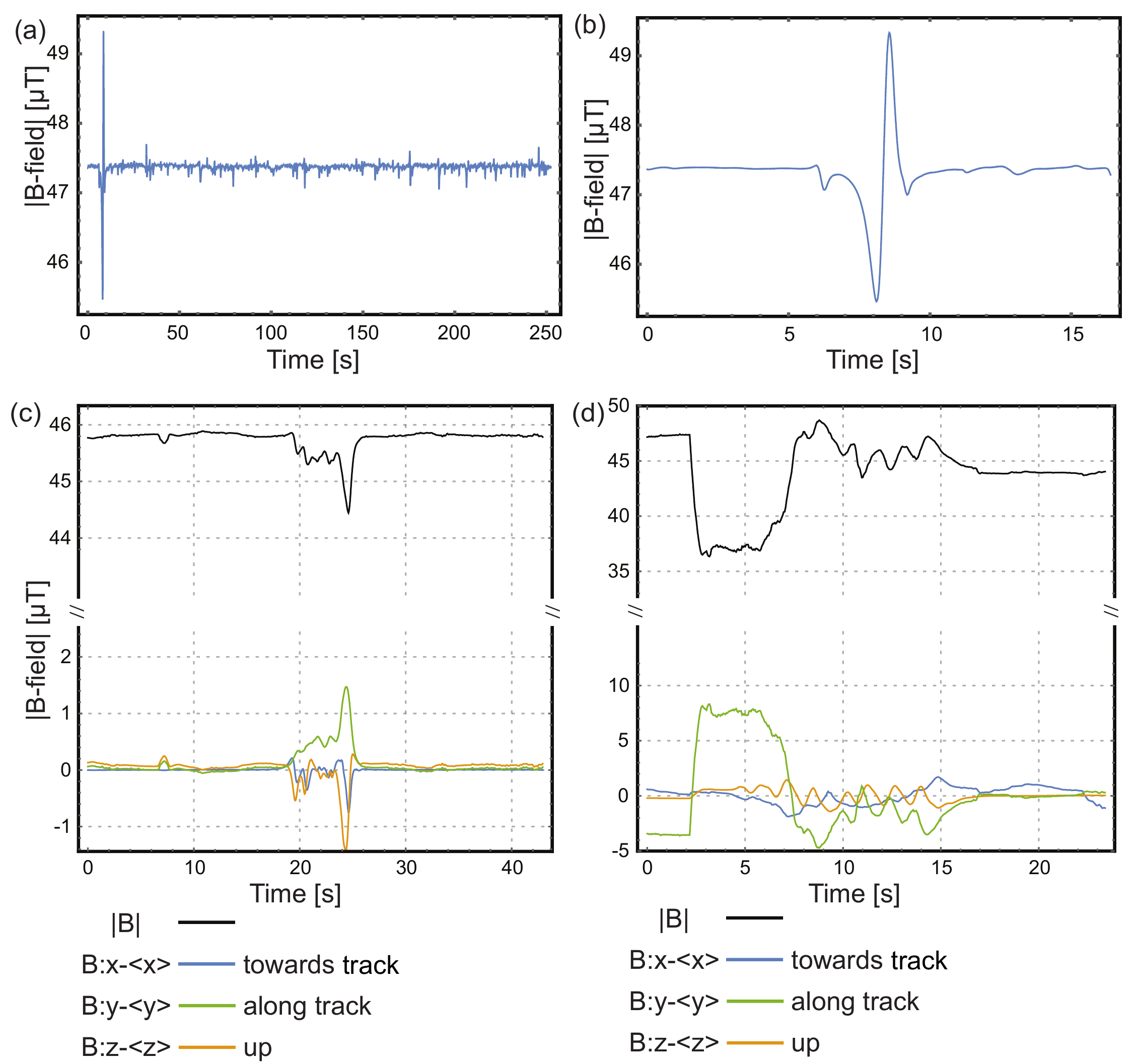}
  \caption{Examples of urban magnetic sources. (a) Scalar magnitude field measurements from sensor placed 5\,m from highway. (b) Signature of large truck passing on highway [first peak in (a)]. (c) Magnetometer placed close to the tracks of a passing Amtrak train. The scalar magnitude field is shown in black with the vector components (DC removed) displayed in color. (d) Data taken at El Cerrito BART station on the side of southbound trains.}
  \label{Figure4}
\end{figure*}

\begin{figure*}[ht]
  \centering
  \includegraphics[width=\textwidth]{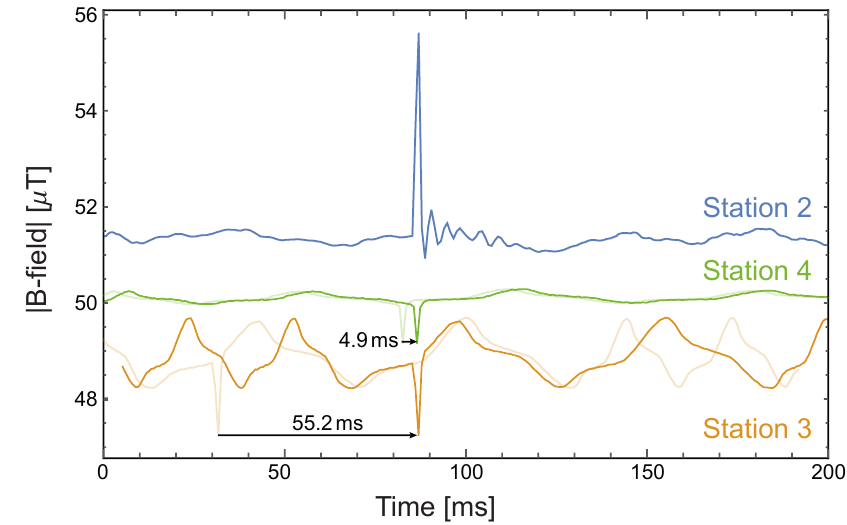}
  \caption{Magnetic anomaly detection. Observations from three geographically distributed magnetometers of a single lighting strike. At the time of the strike, the timing issues described in Section \ref{timing} had not been resolved.  Magnetometer traces from each station were shifted to align the spikes, demonstrating the need for a precision timing algorithm. Looking at the apparent frequency change in the 60\,Hz power-line signature of the yellow trace, we infer that timing errors occurred in this sensor just before the lightning strike. The blue trace was shifted up by 2\,$\mu$T for plotting purposes. Station 1 served as an engineering unit and was not part of the ongoing observations.}
  \label{Figure5}
\end{figure*}

The portability of our sensors has enabled the direct measurement of several urban field sources. Figure \ref{Figure4} shows the magnetic signatures of several field sources associated with transportation: traffic on a freeway, as well as both Amtrak and BART trains. Figure \ref{Figure5} shows a spike in the magnetic field due to a lightning strike recorded by three geographically distributed sensors. This lightning strike, observed before the implementation of the timing algorithm, highlights the need for time corrected data: e.g. interpreting the time lag between sensors as a physical transit time between stations corresponds to unreasonable spatial separations of ${\sim}15,000$\,km. Unfortunately, no further lightning strikes have been captured since implementation of the timing algorithm: the synchronous occurrence of lightning in our network will eventually provide an important test of our timing algorithm.

\subsection{Multi-station analysis of magnetic field data}
\label{multistation}

Fraser-Smith et al. (1978) report the presence of strong ultralow frequency (ULF) magnetic fluctuations throughout the San Francisco Bay Area in the 0.022 to 5\,Hz range \cite{fraser-smith:1978}. The dependence of these fluctuations on proximity to the BART lines, and their correspondence with train timetable, led the authors to attribute this ULF signal to currents in the BART rail system. Subsequent ULF measurements made in \cite{ho:1979} at a distance of ~100\,m of BART suggested periodic bursts of magnetic field at roughly the periodicity of the BART train. The location of our network stations (up to 2\,kms from BART line) and length of recorded intervals ensure that urban signatures, such as BART, will be present in our data.\\

The timing algorithm provides sub-millisecond resolution for MAD; however many magnetic signatures related to anthropogenic activity occur at low frequencies where GPS alone provides adequate timing. To investigate urban magnetic fluctuations, we decimate the full 3,960 sample/s to a 1 sample/s cadence. Antialiasing is accomplished with a moving average (boxcar) filter. Low-cadence data provide adequate resolution for correlating multi-station observations, while simultaneously removing the 60\,Hz power signal. We only use data from Stations 2, 3 and 4 in multi-station comparisons. Station 1 served mainly as an engineering unit for characterisation measurements and other testing.\\

\begin{figure*}[p]
  \centering
  \includegraphics[width=0.85\textwidth]{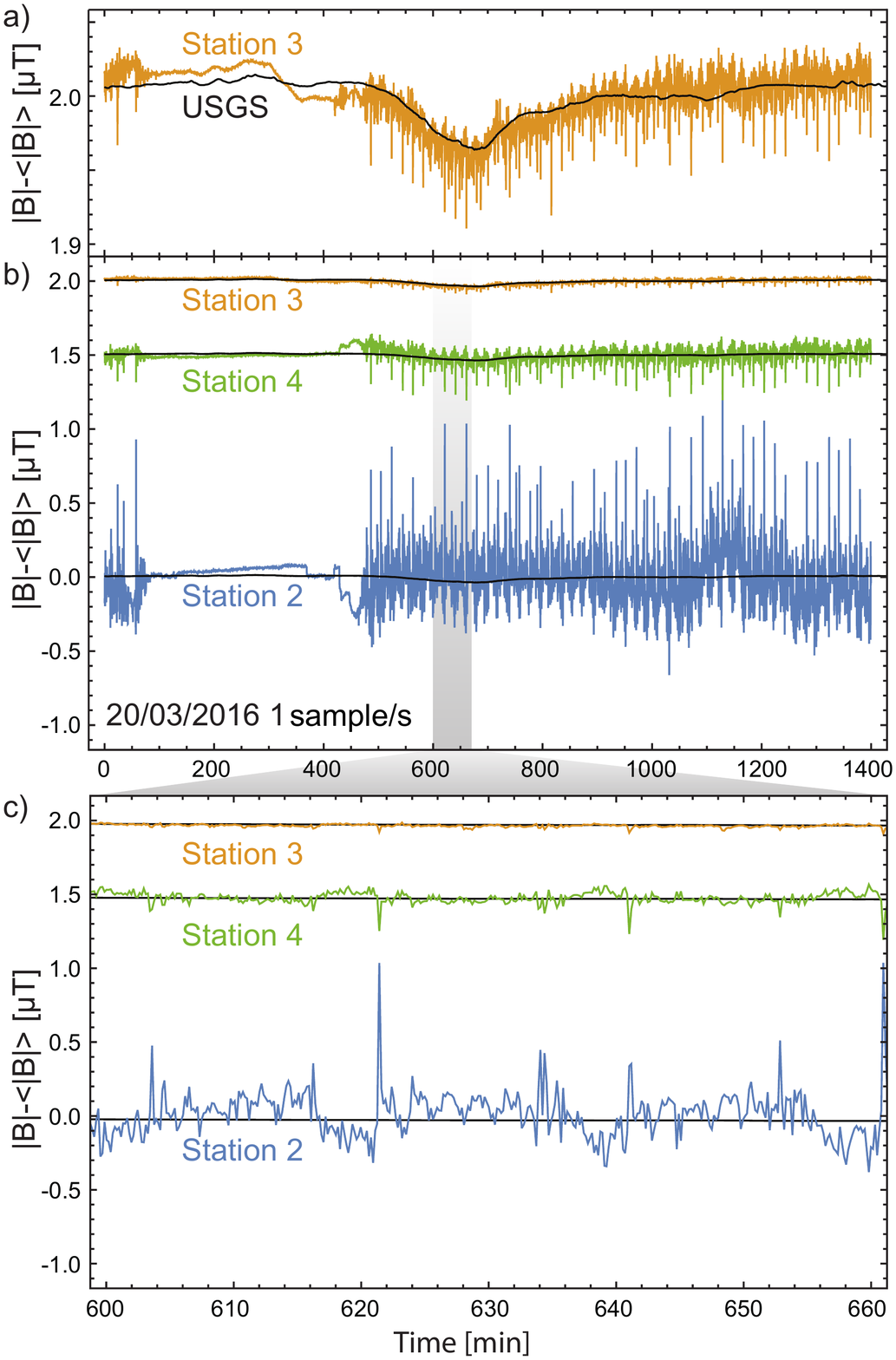}
  \caption{Magnetic field magnitudes for three stations at 1 sample/s cadence on 03-20-2016. Panel (a) shows close agreement between the USGS station in Fresno and sensor 3, demonstrating our stations' sensitivity to geomagnetic fields. Panel (b) shows the 24 hour scalar magnitudes (DC values were adjusted for plotting purposes) for all three active sensors; the magnitude of fluctuations decreases with increasing distance from each sensor to the BART train line. The grey bar represents one hour of data from 10-11\,AM (PDT), a better visualisation of that region can be seen in panel (c). The 24-hour scalar field averages have been removed from each time-series. The one-minute averaged USGS geomagnetic field data are shown in each plot.}
  \label{Figure6}
\end{figure*}

Figure \ref{Figure6} shows the scalar magnitude fluctuations of three stations on Sunday, 03/20/2016 (PDT). One-day scalar average magnitudes are subtracted from the total magnitude. Each panel additionally shows geomagnetic field data (one-minute averaged) acquired from the United States Geological Survey (USGS) station in Fresno, CA. Though panel (a) demonstrates a general agreement between our record and the USGS data, consistent large fluctuations in excess of the geomagnetic field are observed in each sensor. Panel (b) shows that these non-geomagnetic fluctuations dominate the daytime magnetic field. Additionally, panel (c) shows a subset of the data from 10-11\,AM, revealing several synchronous spikes in each sensor.\\

\begin{figure*}[ht]
  \centering
  \includegraphics[width=0.75\textwidth]{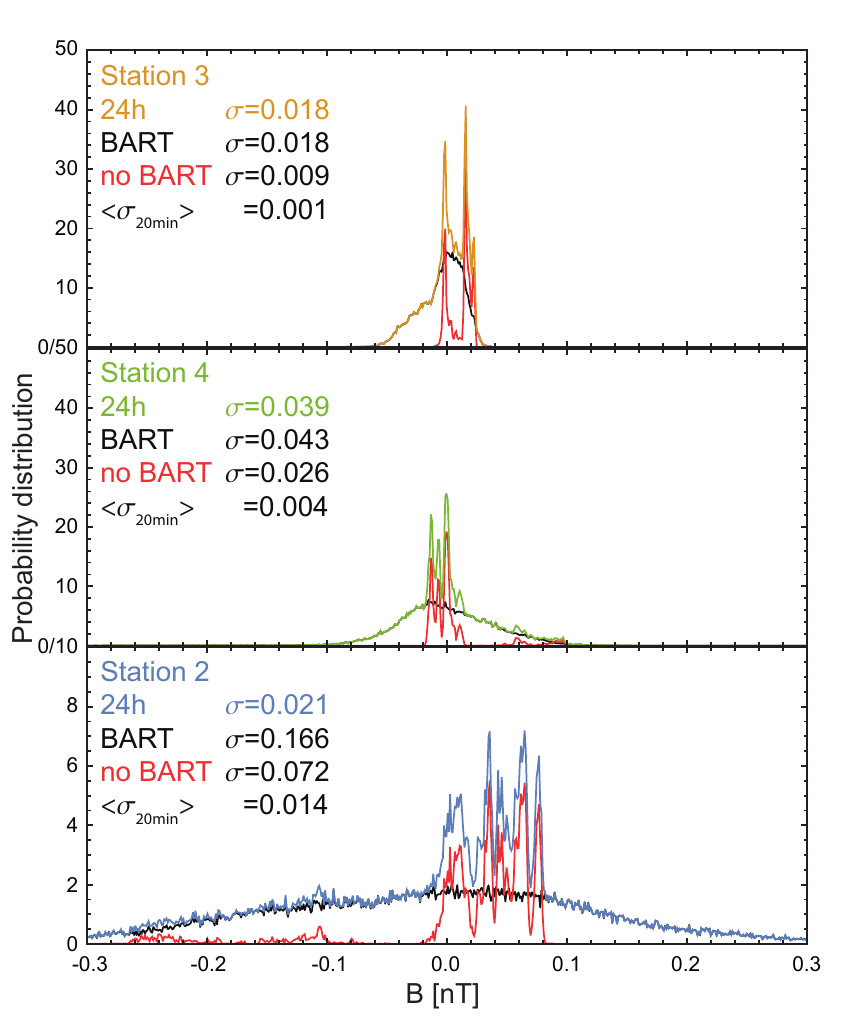}
  \caption{Distributions for 24hr, daytime, and nighttime magnetic field observations. The nighttime observations show several discrete peaks, daytime fluctuations follow broad distributions. The variance of the distributions increases as the distance from the BART train line decreases.}
  \label{Figure7}
\end{figure*}

A magnetically quiet period corresponding to BART non-operating hours \footnote{\url{https://www.bart.gov/}} is evident in the records of the three active sensors (stations 2-4). Figure \ref{Figure7} shows the distribution of magnitude fluctuations binned at 10$^{-3}$\,nT. The record naturally separates into two intervals, corresponding to the hours when BART trains are running (roughly, from 7:55AM to 1:26AM) and hours when BART trains are inactive. We refer to these intervals as ``daytime'' and ``nighttime'', respectively.  The magnetic measurements reveal characteristically different distribution functions in daytime and nighttime. For each sensor, the nighttime distribution functions appear as a superposition of several individual peaks. Standard deviations, $\sigma_i$, for the nighttime distribution functions of the three active sensors are given by $\sigma_{2}=0.072\,\mu$T, $\sigma_{3}=0.009\,\mu$T, and $\sigma_{4}=0.026\,\mu$T.\\

\begin{figure*}[ht]
  \centering
  \includegraphics[width=\textwidth]{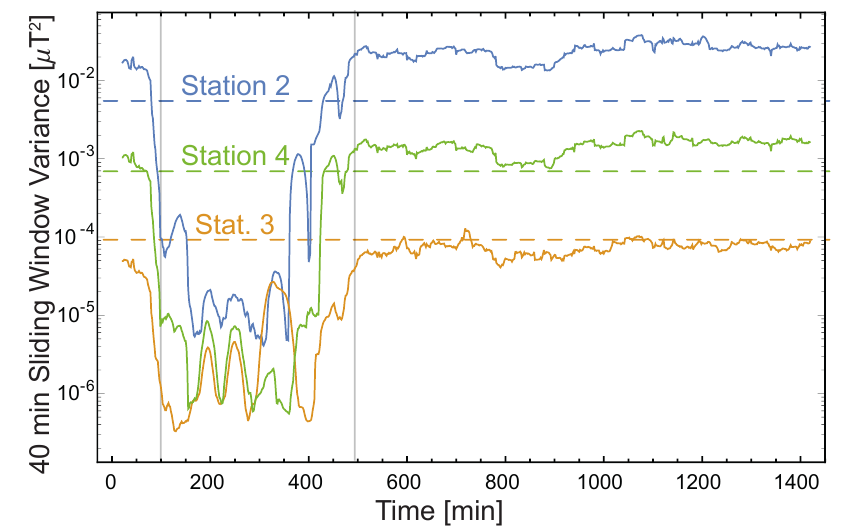}
  \caption{Magnetic field variance calculated over a 40-minute sliding window for all three sensors. Vertical lines mark the night interval when BART is inactive.  Strong, discontinuous, transitions in the DC field amplitude, observed as peaks in the sliding window variance, cause the variance calculated over the entire night interval to be significantly larger than the time-local variance.}
  \label{Figure8}
\end{figure*}

Figure \ref{Figure8} shows that the time-localized variance of magnetic field fluctuations, calculated in a 40 minute sliding-window, is significantly smaller than the variance calculated for the full nighttime interval. This indicates that the discrete peaks observed in the nighttime distribution functions are localized in time, while transitions in the DC field magnitude cause the appearance of several distinct peaks. These transitions in the DC field are evident as peaks in the sliding-window variance. The simultaneous occurrence of transitions in the DC field magnitude, evidenced by simultaneous peaks in the sliding window variance, suggests that these transitions are global phenomena, perhaps relating to nighttime maintenance on the BART line or activity in the ionosphere.\\

 The two vertical bars in Fig. \ref{Figure8} mark the period for which BART is inactive. Increased field variance, i.e. signal power, is clearly tied to the operation of BART. Figure \ref{Figure8} also highlights the relatively constant daytime variance in each station. Additionally, there appear to be approximately constant ratios between the daytime variance (power) observed by each sensor. This suggests that the background noise level observed in each station is set by the distance to BART.  Identifying other anthropogenic fields (for example, traffic) is complicated by the large BART signal. Accordingly, identification of further urban signals requires a thorough characterization of the magnetic background generated by BART.

 \subsection{Time-Frequency (Wavelet) Analysis}
 \label{correlation}

Frequency-domain analysis is typically used to reveal the spectral composition of a magnetic time series. Localizing the distribution of spectral power in time requires simultaneous analysis in both time and frequency domains. We implement a continuous wavelet transform (CWT), using Morlet wavelets \cite{Torrence:1998}. to investigate the time-frequency distribution of low-frequency fluctuations associated with BART. The unnormalized Morlet wavelet function $\psi(\tau)$ is a Gaussian modulated complex exponential,

\begin{equation}
\psi (\tau)= \pi^{1/4}e^{i\omega_0\tau }e^\frac{-\tau^2}{2},
\end{equation}

\noindent with non-dimensional time and frequency parameters $\tau$ and $\omega_0$. A value of $\omega_0=6$ meets the admissibility conditions prescribed in Ref.\ \cite{Farge:1992} and is commonly used across disciplines \cite{Podesta:2009,Torrence:1998}. At each time step, the CWT of the magnetic field $B$ is defined by the convolution of the time series record with a set of scaled wavelets,

\begin{equation}
W(s,t)=\sum_{i=0}^{N-1}B_x(t_i)\psi(\frac{t_i-t}{s}),
\end{equation}

which are normalized to maintain unit energy at each scale. The CWT provides a scale independent analysis of time-localized signals, and is additionally insensitive to time series with variable averages (non-stationary signals). These qualities provide some advantage over alternative time-frequency analysis techniques, such as the windowed Fourier transform, which calculates the spectral power density in a sliding window applied to the time series. Introducing a window imposes a preferred scale which can complicate analysis of a signal's spectral composition. For example, low-frequency components, with periods longer than the sliding window scale, are aliased into the range of frequencies allowed by the window, thereby degrading the estimate of spectral density.\\

\begin{figure*}[ht]
  \centering
  \includegraphics[width=0.85\textwidth]{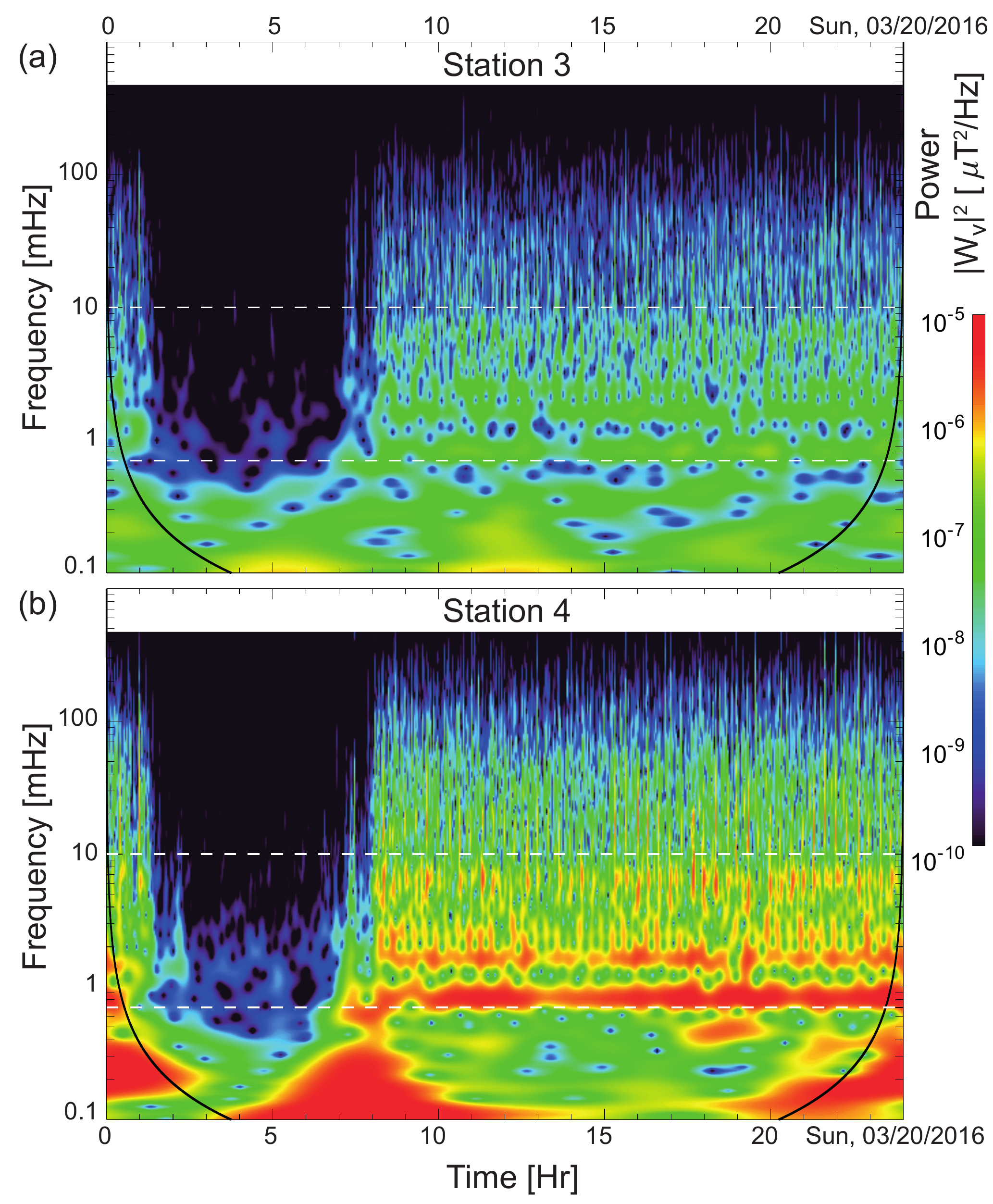}
  \caption{Time-frequency analysis of scalar magnitude magnetometer data. Continuous wavelet transform power spectral densities ($\mu$T$^2$/Hz) for Magnetometer 3 (a) and Magnetometer 4 (b).  The spectrograms reveal power in several bands from $8.3\times10^{-4}$  and $1\times10^{-2}$\,Hz common to both sensors. These frequencies correspond to a 20-minute period signal and subsequent harmonics. White dashed lines show the frequency range of a brick-wall filter applied to data.}
  \label{Figure9}
\end{figure*}

Full day, 1 sample/s cadence, wavelet power spectral densities $|W(s,t)|^2$  ($\mu\textrm{T}^2/\textrm{Hz})$ for stations 3 and 4 on Sunday, 03/20/2016 (PST) are displayed in Fig. \ref{Figure9}. These spectrograms prominently display the quiet nighttime period. Additionally, strong power is observed in several scales corresponding to a fundamental 20-minute period ($8.33 \times 10^{-4}\ \textrm{Hz}$) and associated higher harmonics. This 20-minute period coincides with the Sunday BART timetable on the geographically closest BART line (Richmond-Fremont). The black lines display the region where boundary effects are likely\ --\ this region, known as the cone of influence (COI), corresponds to the $e$-folding time for the wavelet response to an impulse function. \\

\begin{figure*}[ht]
  \centering
  \includegraphics[width=0.73\textwidth]{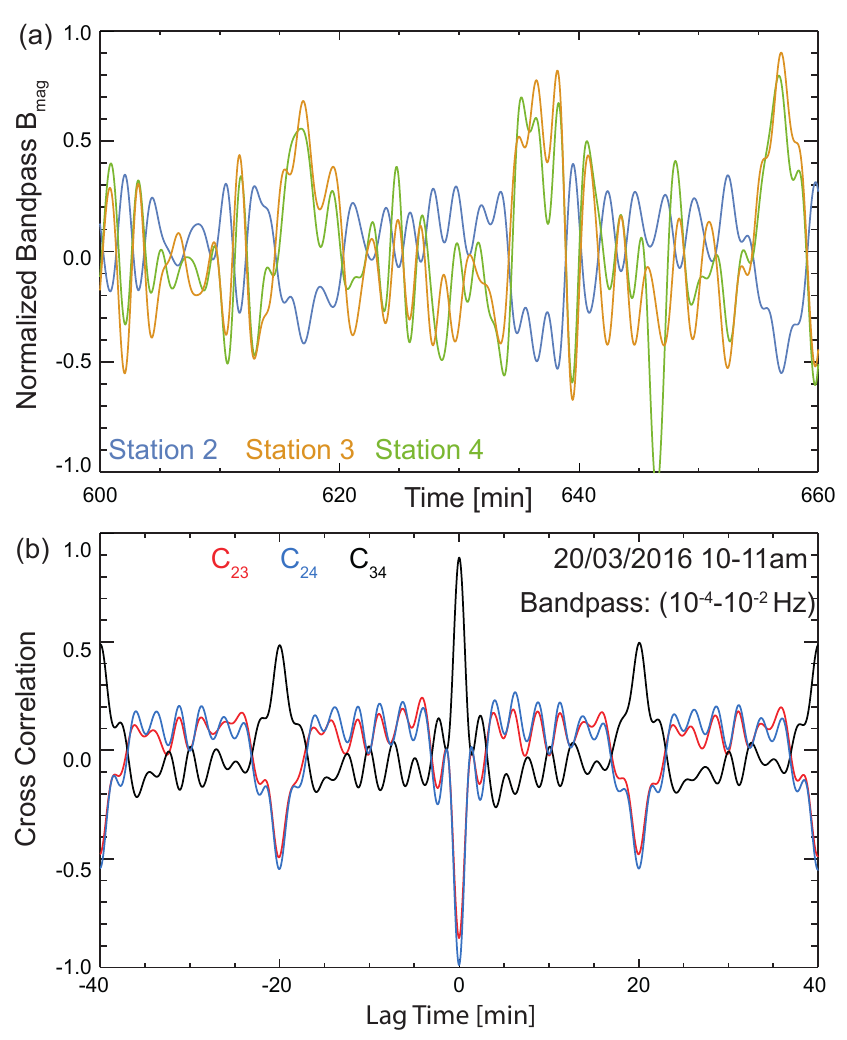}
  \caption{(a) Brickwall bandpassed ($7 \times 10^{-4}$  to $1 \times 10^{-2}$\,Hz) time-series for all three sensors on 03-20-2016 between 10-11\,AM (PDT). The bottom panel shows the correlation coefficients between pairs of sensors as a function of lag. Sensors 3 and 4 are highly correlated (in phase), while sensor 2 is anti-correlated (out of phase) with the others. There is a 20-minute periodicity to the data, consistent with the powerbands observed in the wavelet spectra. This 20-minute signal coincides with the published Sunday/Holiday BART schedule for the Fremont/Richmond train line.}
  \label{Figure10}
\end{figure*}

A brick-wall bandpass filter (i.e., unity gain in the passband, full attenuation in the stop band) is applied to each sensor in the frequency domain between $7 \times 10^{-4}$  and $1 \times 10^{-2}\ \textrm{Hz}$ in order to isolate the bands of power observed in the wavelet power spectra. The top panel of Fig. \ref{Figure10} shows the bandpassed time series for 10-11\,AM on 3/20/2016. The bandpassed time series are normalized to their maximum values for the purpose of visualization. This plot immediately suggests that stations 3 and 4 are highly correlated, while station 2 is anti-correlated with stations 3 and 4. Indeed, this is verified by the bottom panel of Fig. \ref{Figure10}, which shows the cross correlation coefficients $C_{ij}(\tau)$ calculated for the 24 hour time series:

\begin{equation}
C_{ij}(\tau)=\frac{\sum_{n=0}^{N-\tau-1}(B_i[n+\tau]-\bar{B_i})(B_j[n]-\bar{B_j})}{\sqrt{\left[{\sum_{n=0}^{N-1}(B_i[n]-\bar{B_i})^2}\right]\left[{\sum_{n=0}^{N-1} (B_j[n]-\bar{B_j})^2}\right]}},
\end{equation}

\noindent where $N$ is the record length, $\tau$ is a translation between time series, and $n$ is the sample index \cite{Bendat:1990}. It is clear that stations 3 and 4 are in phase, while station 2 is out of phase of the other two instruments. These phase relationships which correspond to the geographical location of the sensors on the east/west side of the BART line (stations 3 and 4 are located east of the rails, while station 2 is located to the west, c.f. Fig. \ref{map}), suggest that the magnetic field generated by the BART has a strong azimuthal symmetry around the BART rail. Future work will look to determine the multipole components (e.g. line current, dipole, and higher corresponding to the BART field.\\

\begin{figure*}[ht]
  \centering
  \includegraphics[width=0.85\textwidth]{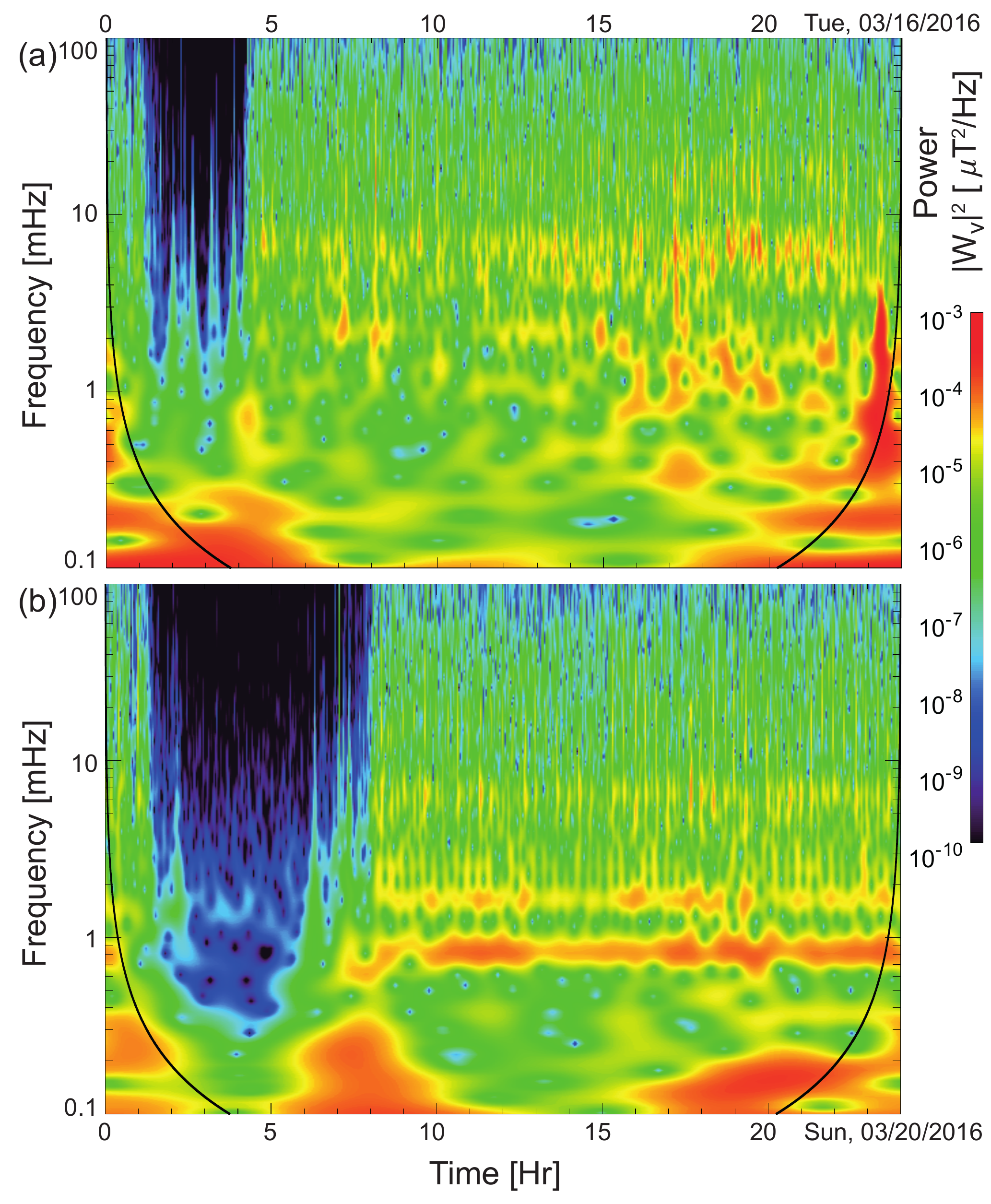}
  \caption{BART night signatures. (a) Continuous wavelet transform of 1 sample/s magnetic field magnitude data from Station 2 on 3/16/2016 (and 3/20/2016). The nighttime signature is significantly shorter in the data taken on Wednesday, this corresponds with BART operating hours. Additionally, the strong powerbands observed on Sunday are not present in the Wednesday data.}
    \label{Figure11}
\end{figure*}

Figure \ref{Figure11} shows full day wavelet spectral densities for station 2 on both Wednesday, 03/16/2016 and Sunday, 03/20/2016. The quiet BART night is much shorter on Wednesday; this corresponds with the different weekend and weekday timetables. Additionally, Fig. \ref{Figure11} demonstrates the absence of a strong 20-minute period in the Wednesday data, instead revealing more complex spectral signatures. The increase in complexity observed in the weekday spectrogram is most certainly due to increases in train frequency, the addition of another active BART train, and variability associated with commuter hours. In our future work we will fully explore the effect of train variability on the urban magnetic field using correlated observations from the entire network.\\

The measurements of \cite{ho:1979} made between 0.001 and 4 Hz, at a range of several hundred meters from the BART rail, captured bursts of magnetic field corresponding approximately to the train schedule, but with an irregular variation in the waveform of the magnetic field. In contrast, our observations reveal a highly regular signature, with multiple spectral components, occurring at the BART train period. The repeatability of this signature, observed coherently in the three deployed sensors, enables for identification and extraction of the periodic waveform associated with the BART operation. 

\section{Extracting the BART signal}
\label{extraction}

\begin{figure*}[ht]
  \centering
  \includegraphics[width=0.73\textwidth]{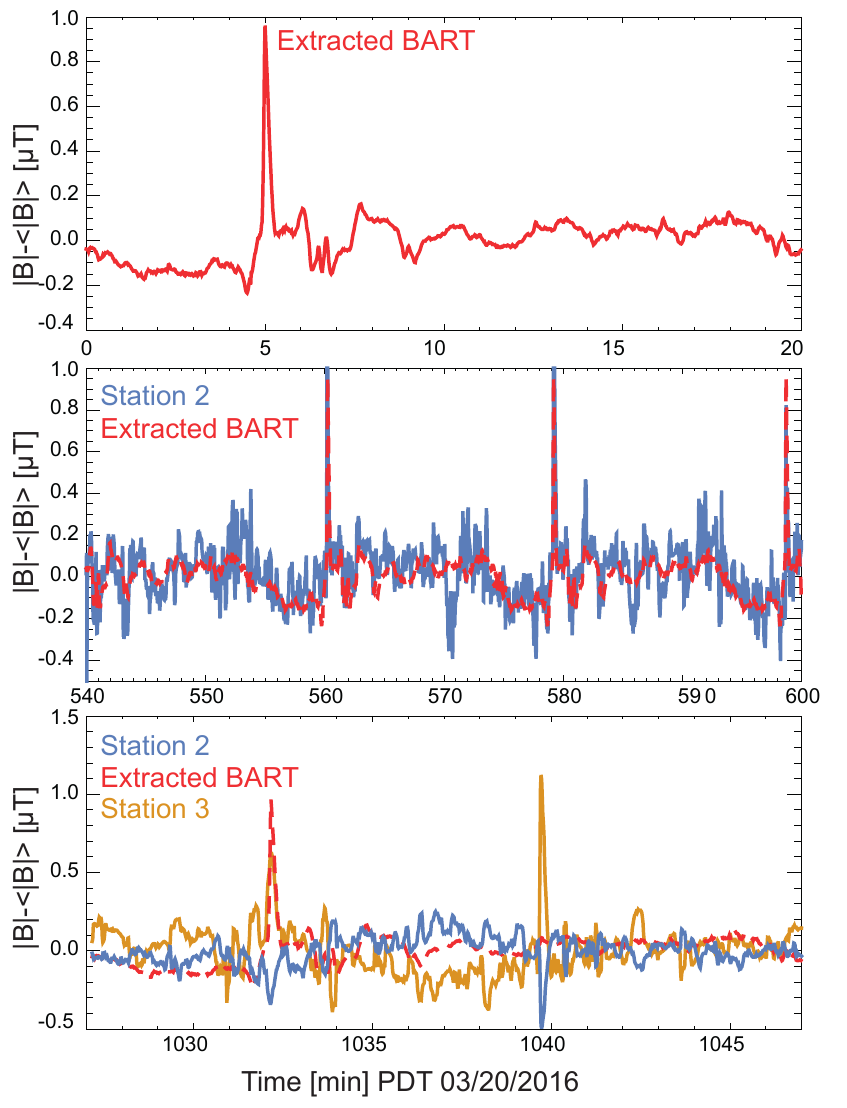}
  \caption{Extraction of BART signal. (a) 20-minute periodic signal of the BART extracted from an ensemble average of 46 intervals from station 2. (b) Comparison of the extracted average signal with an hour of observations from station 2 taken from 9-10 AM (PDT).  (c) Comparison of extracted average signal with an interval containing local magnetic anomaly, (d) Comparison of extracted average signal with an interval containing global magnetic anomaly likely due to variation in BART operation.}
    \label{bartextract}
\end{figure*}

Liu \& Fraser Smith \cite{liu:1998} attempt to extract features associated with the BART using wavelets to identify transient features in a geomagnetic time-series. Our observations of a periodic BART signature enable a statistical averaging over the observed period in order to extract features related to BART. The periodic 20-minute signature observed in the Sunday 03/20/2016 time-series from station 2 can be extracted using the technique known as superposed epoch analysis \cite{singh:2006}. We identified 46 sharp peaks in the magnetic field occurring with an approximately 20-minute period (e.g. Fig. \ref{Figure6}). From these 46 peaks, an ensemble $[X(t)_i]$ of intervals is constructed comprising the 3 minutes preceding and 17 minutes succeeding each individual peak. Averaging over the ensemble of intervals $\bar{X}(t)=\sum_i X_i(t)$ reveals a coherent signature with an approximate 20-minute period, Fig. \ref{bartextract}(a). The periodic signal observed in the data has the form of a sharp discrete peak of $\approx$1\,nT, followed by an oscillation with a period on order of several minutes.  A quantitative comparison is obtained through computing the Pearson correlation

\begin{equation}
  \rho_i=\frac{{cov}(X_i,\bar{X})}{\sigma_{X_i}\sigma_{\bar{X}}}
\end{equation}

of the extracted signal with each interval in the ensemble. On average, the correlation between the extracted signal and observed data has a correlation of $\bar{\rho}=0.7$, with $\rho_i$ ranging from 0.1 to 0.85. We can interpret these values as the fraction of power in each interval derived from the average signature. Figure \ref{bartextract}(b) demonstrates high correlation between the extracted average signal with an hour of observations taken from 9-10\,AM (PDT). Through extracting periodic magnetic signatures of BART, we enable the identification of transient events associated with BART operation as well as other urban phenomena. Fig.\ref{bartextract}(c) shows  the occurrence of a local magnetic anomaly occurring at 12PM (PDT) in Station 2; in this case $\rho=0.67.$ The traces from the other stations, suggest that the event observed in station 2 is a local anomaly, not associated with BART. Additionally, Figure \ref{bartextract}(d) demonstrates an interval of data (with $\rho=0.17$) which includes a global transient feature, likely due to some variation in the BART system. Measurements from a single sensor allow us to identify events which deviate from the correlated periodic observations; our future work will employ the full network of magnetometers to identify correlated signals in both space and time, allowing for an extraction of the magnetic field local to each sensor from the global field dominated by the BART signal.

\section{Discussion}
\label{discussion}

An array of four magnetometers has been developed with bandwidth of DC-kHz and sensitivity better than 0.1\,nT/$\sqrt{\textrm{Hz}}$. The array is currently deployed in the area surrounding Berkeley, CA, providing measurements of an urban magnetic field. This array is sensitive to both natural magnetic activity, such as lightning and the low frequency variations in the Earth's geomagnetic field, as well as a variety of anthropogenic sources: currents associated with BART, traffic and 60 Hz powerlines. The operation of BART dominates the urban magnetic field generated broadband noise. In addition to this broadband noise, the network has identified the presence of coherent narrowband spectral features originating from the BART. Significant variation in the spectral features is observed between weekends and weekdays corresponding to variations in the BART train schedule. During the hours in which BART is non-operational, the anthropogenically generated fields are significantly decreased and agreement with the USGS magnetic field measurements is observed. However, the nighttime field still contains a number of features not attributable to geophysical activity. Further study is required to determine the nature and sources of these features.\\

Cross-correlating the sensors at high frequencies requires a high-precision timing algorithm to combine the absolute time, acquired through GPS, with the high-precision computer performance clock local to each station. This algorithm additionally corrects for latency issues associated with the USB interface between the data-acquisition hardware and the laptop operating system. This timing algorithm has been tested using magnetic fields generated by Helmholtz coils. We intend to use the impulsive globally-observable fields generated by lightning to further test our timing algorithm. Our high precision timing will allow for such magnetic anomaly detection on the order of $\approx100$\uS.\\

This paper presented a proof-of-concept deployment of what, to our knowledge, is the first synchronized network of magnetometers specifically designed for observing the effects of human activity on the magnetic field in an urban environment. Numerous potential applications and directions for future work have emerged. Further development of algorithms to remove the BART signal (or any other dominant signal whose source has been identified)  must be developed. These algorithms may need to take advantage of other data sources (e.g., the realtime BART schedule) and machine learning techniques. The study of high frequency response (60\,Hz and above) has not yet been pursued.  We note that anthropogenic fields mask geophysical field fluctuations, and that study of the latter is facilitated by understanding of anthropogenic noise.  The magnetic fields due to humans may reflect identifiable aspects of urban dynamics (beyond BART) and these may have correlations to other measures of urban life (energy consumption being one of the first to consider). Studies of magnetic field correlations and anomalies may be used to identify and study local phenomena (traffic, elevators, etc.). One spin-off of this research may be improved identification and reduction of anthropogenic noise in geomagnetic measurements located in or near urban environments. The ultimate utility of the magnetometer array as an observational platform for urban systems will only become clear with further studies. 

\section{Acknowledgements}

We are grateful to Brian Patton for his contributions in the early stages of the project. The views expressed in the publication are the author's and do not imply endorsement by the Department of Defense or the National Geospatial-Intelligence Agency.

\bibliography{references}

\end{document}